\documentclass[twocolumn,
nofootinbib,
amsmath,amssymb]{revtex4-1}

\usepackage{hyperref}
\usepackage{dsfont}
\usepackage{overpic}

\newcommand{\nn}{\nonumber}

\def\be{\begin{equation}}
\def\ee{\end{equation}}
\def\bea{\begin{eqnarray}}
\def\eea{\end{eqnarray}}
\def\bal{\begin{align}}
\def\eal{\end{align}}
\def\zb{\bar{z}}
\def\ub{\bar{u}}
\def\vb{\bar{v}}
\def\Im{\textrm{Im}}
\def\Re{\textrm{Re}}
\def\K{\mathcal{K}}
\def\W{\mathcal{W}}

\begin{document}
\title{Winding out of the Swamp:\\[.2cm] 
Evading the Weak Gravity Conjecture with $F$-term Winding Inflation?}

\author{Arthur Hebecker}
\email{A.Hebecker@ThPhys.Uni-Heidelberg.de}
\author{Patrick Mangat}
\email{P.Mangat@ThPhys.Uni-Heidelberg.de}
\author{Fabrizio Rompineve}
\email{F.Rompineve@ThPhys.Uni-Heidelberg.de}
\author{Lukas T.\ Witkowski}
\email{L.Witkowski@ThPhys.Uni-Heidelberg.de}
\affiliation{Institute for Theoretical Physics, University of Heidelberg, Philosophenweg 19, 69120 Heidelberg, Germany}

\begin{abstract}
We present a new model of large field inflation along a winding trajectory in the field space of two axionic fields, where the ``axions'' originate from the complex structure moduli sector of a Calabi-Yau 3-fold at large complex structure. The winding trajectory arises from fixing one combination of axions by bulk fluxes and allows for a transplanckian effective field range. The inflaton potential arises from small ``instantonic'' corrections to the geometry and realises natural inflation. By working in a regime of large complex structure for two complex structure moduli the inflaton potential can be made subdominant without severe tuning. We also discuss the impact of the recent `no-go theorems' for transplanckian axion periodicities on our work. Interestingly, our setup seems to realise a loophole pointed out in arXiv:1503.00795 and arXiv:1503.04783: our construction is a candidate for a string theory model of large field inflation which is consistent with the mild form of the weak gravity conjecture for axions.
\end{abstract}

\maketitle 

\section{Introduction}
The question whether large field inflation can be realised in string theory is one of the great open questions of string phenomenology. On the one hand, many ideas for how large field inflation could be embedded in string theory have been proposed, while on the other hand, there exist arguments that Quantum Gravity should censor any attempt at constructing such models \cite{0601001, 12035476}. Progress can be made from two sides: for one, by sharpening no-go theorems in string theory one can reduce the available `model space' for large field inflation. Alternatively, by constructing a successful model one can hope to arrive at a proof by existence. Here we follow the latter course and propose a new model of large field inflation in string theory which evades some current no-go theorems.

There are various avenues towards large field inflation in string theory. One approach employs the fact that a transplanckian field range can arise in the field space of two or more axions \cite{0409138, 0507205}. Long directions can be constructed along a diagonal of the field space or they can arise if the inflaton trajectory winds around the compact space many times before returning to the starting point (see in particular \cite{09121341}). In the latter case the technical difficulty is to construct a potential that forces the inflaton onto this trajectory. Another proposal for large field inflation in string theory employs monodromies which extend the originally compact field space of an axion \cite{08033085, 08080706, 08111989}. Recently, a wealth of models based on all these approaches appeared in the literature \cite{14037507, 14043040, 14043542, 14043711, 14044268, 14045235, 14046209, 14046988, 14047127, 14047496, 14047773, 14047852, 14050283, 14052325, 14053652, 14060341, 14072562, 14097075, 14104660, 14112032, 14114768, 14115380, 14121093, 150301607, 150301015, 150302965, 150307183, 150307634}. 

However, models based on realising long trajectories in the field space of two or more axions are currently under pressure \cite{14095793, 14123457, 150300795, 150303886, 150304783}. In \cite{150303886} it is pointed out that gravitational and string instantons tend to spoil transplanckian inflaton trajectories, while in \cite{150304783} it is argued that transplanckian axion periodicities are inconsistent with (the strong form of) the weak gravity conjecture \cite{0601001}. 

These arguments do not apply immediately to models based on monodromies, which unfortunately have their own problems. The challenges for such axion monodromy constructions can be studied very explicitly in the framework of $F$-term axion monodromy models \cite{14043040, 14043542, 14043711}, which exhibit increased qualitative and quantitative control compared to previous proposals. In this context it was found that moduli stabilisation constraints forbid the universal axion as an inflaton \cite{14097075}. In addition, $F$-term axion monodromy models of quadratic inflation in the complex structure moduli sector of Calabi-Yau 3/4-folds have to be severely tuned \cite{14112032}. It is also expected that supersymmetry breaking effects cannot be ignored in these models \cite{14070253, 150105812}.

However, inflation in the complex structure moduli sector is not restricted to models with monomial inflaton potentials. The construction of natural inflation in this moduli sector has been attempted in \cite{14114768}. Here, in the spirit of \cite{0409138, 09121341}, we propose a model which achieves a large field displacement by realising a winding trajectory in the compact field space of two complex structure ``axions''. The model has two advantages. For one, the inflaton potential arises from ``instantonic'' terms and is sufficiently small not to interfere with moduli stabilisation without severe tuning. In addition, by realising a loophole pointed out in \cite{150300795, 150304783} it has the potential to evade the recent no-go theorems. Note that a different loophole exists \cite{14123457} (see also \cite{150300795}). We discuss it in Sec.~\ref{sec:nogo}.

The model is based on three observations. 
\begin{itemize}
\item In the limit of large complex structure, complex structure moduli exhibit axionic shift symmetries.\footnote{For shift symmetries at other special points in complex structure moduli space and possible applications to inflation see \cite{14125537}.}
\item Bulk fluxes give rise to an $F$-term potential for these axions. By flux choice one can fix some of the axions such that only one flat axionic direction remains \cite{14097075, 14114768}. By choosing fluxes appropriately one can ensure that the flat direction is a long, winding trajectory which wraps the compact field space multiple times (see e.g. figure \ref{fig:trajectory}).  
\item Corrections to the large-complex-structure geometry (corresponding to instantons in the mirror dual 3-fold) generate a potential along the previously flat direction. There are two types of relevant instantonic terms: those with a long period, corresponding to the full length of the trajectory, and those with a short period, potentially ruining inflation. We stabilise the saxions such that the long-period terms are reasonably small while the dangerous terms are completely negligible. 
\end{itemize}
In particular, our inflationary direction is similar to the one constructed in \cite{14047773}. In both cases, one direction in the field space of two axions is fixed by a dominant potential term. However, our scenario differs from previous proposals, and in particular from \cite{14047773}, for the following reasons.
Firstly, in our model the winding trajectory is enforced by an $F$-term potential from bulk fluxes. This is closely related to the approach taken in \cite{150302965} where the long field range is realised via brane fluxes, by gauging away certain combinations of axions. In contrast to our model, in \cite{14047773} the winding trajectory is enforced by non-perturbative contributions to the superpotential.
Secondly, as we discuss in Sec.~\ref{sec:nogo}, our model realises the loophole pointed out in \cite{150300795, 150304783}. The mechanism proposed in \cite{14047773} requires an axion potential with two non-perturbative contributions. One such term is responsible for generating the winding trajectory, while the other will give rise to the inflaton potential. As we discuss in Sec.~\ref{sec:nogo}, this minimal setup of \cite{14047773} does not evade the recent no-go arguments based on the WGC.

Interestingly, our construction comes with a cutoff to the possible field range for the inflaton. The long winding trajectory in the field space of two complex structure axions can be constructed at the expense of choosing large values for some of the flux numbers. However, tadpole cancellation will lead to a bound on the allowed values for flux quanta such that arbitrarily large axion periods are prohibited. 

In addition, we wish to stress that the model crucially relies on the pattern of moduli stabilisation: while the axions are stabilised such as to realise the winding trajectory, the saxions need to be fixed at large complex structure. The model could be excluded if this required scheme of moduli stabilisation was shown to fail. While we do not see any fundamental obstructions, this point requires further scrutiny.

The resulting inflaton potential realises natural inflation and is thus potentially in conflict with the recent ``no-go theorems'' in \cite{150303886, 150304783}. We will discuss the impact of this work on our model in more detail in the text. Here, we would like to point out that while these recent analyses are exciting and important, further work needs to be done to extend these no-go theorems to more realistic string models including further branes, orientifold planes and fluxes.  

Nevertheless, we can make an important observation. In \cite{150304783} the weak gravity conjecture is used to constrain models of axion inflation in string theory. Assuming that the analysis of \cite{150304783} applies to all models of axion inflation in string theory the results imply the following:
\begin{itemize}
\item If the strong form of the weak gravity conjecture is obeyed by string theory, all string models of inflation with transplanckian axion periods are ruled out. We then expect our model to exhibit pathologies when studied in more detail.
\item If the weak gravity conjecture only applies in its mild form, some models exhibiting axions with a transplanckian field range are still possible \cite{150304783}. Intriguingly, our model appears to fit precisely into this loophole.
\end{itemize}

In the following, we introduce the model, derive the inflaton potential and check that inflation is in principle possible. We then discuss constraints from the no-go results \cite{150303886, 150304783} and explain how this setup is indeed a realisation of the loophole in \cite{150300795, 150304783}.\footnote{After this paper was submitted to the arXiv a related work \cite{150307853} appeared which also critically examines the consequences of the recent no-go results on models of large field inflation in string theory.}

\section{The Model}
Our inflation model is formulated in the effective supergravity description of type IIB string compactifications, where our inflaton arises from the complex structure moduli sector. The model is captured by the following K\"ahler potential and superpotential
\begin{align}
\label{eq:K1} \mathcal{K} &=- \log \left( \mathcal{A}(z, \zb, u-\ub, v-\vb) \right.\\
\nn & \hphantom{=- \log} \left. \ + \left[\mathcal{B}(z, \zb, v-\vb) e^{2 \pi i v} + \textrm{c.c.} \right] \right) \ , \\
\label{eq:W1} \W &= w(z) + f(z) (u - N v) + g(z) e^{2 \pi i v} \ ,
\end{align}
with $N \in \mathbb{Z}$ where we take $N \gg 1$.\footnote{This choice is similar to the one made in \cite{14047773}, which we discuss in more detail in Sec.~\ref{sec:nogo}.} These are the necessary ingredients to compute the F-term scalar potential. The latter is minimised by imposing the conditions $D_{I}\mathcal{W}=0$, where $I$ runs over all moduli. Assuming that the exponential terms in \eqref{eq:K1} and \eqref{eq:W1} are suppressed, only the fields Im$(u)$ and Im$(v)$ as well as the combination Re$(u-Nv)$ are stabilised at the minimum. The fourth scalar component parametrises a flat direction which is closely aligned with Re$(u)$. As will be shown, the exponential terms lift this flat direction introducing a cosine-potential with a large period. 

In the following we will examine in more detail how the structure of \eqref{eq:K1} and \eqref{eq:W1} arises from the geometry of Calabi-Yau 3-folds and argue that the resulting scalar potential is suitable for inflation.

\subsection{Ingredients}
We begin with a type IIB Calabi-Yau orientifold with $h_{-}^{2,1}(X)=n$ complex structure moduli $\{z^i \}$. The quantum-corrected K\"ahler potential can be written as (see e.g.~\cite{9403096, 13023760})
\begin{align}
\nn \K &= - \log \left( - \frac{i}{3!} \kappa_{ijk} (z^{i} - \zb^{i})(z^{j} - \zb^{j})(z^{k} - \zb^{k}) + ic \right. \\
\nn & \hphantom{= AAAa} +i \sum_{\substack{\beta \in H_2(\tilde{X}, \mathbb{Z}) \\ \setminus \{0 \}}} \ \sum_{m=1}^{\infty} \frac{2 n_{\beta}[1-\pi i m \beta_{i}(z^{i}-\zb^{i})]}{(2 \pi i m)^3} \\
 & \hphantom{= AAAAAAAAa} \left. \times \left[e^{2 \pi i m \beta_{i} z^{i}} + e^{-2 \pi i m \beta_{i} \zb^{i}} \right] \vphantom{\frac{1}{3!}} \right) \ ,
\end{align}
where $n_{\beta}$ are constants related to Gromov-Witten invariants, $\kappa_{ijk}$ are intersection numbers of the dual 3-fold and $c =- \frac{i}{4 \pi^3} \zeta(3) \chi(X)$. Summations over $i,j,k$ run from $1$ to $h_{-}^{2,1}(X)$.

The structure for $\mathcal{K}$ in \eqref{eq:K1} can be achieved when working in a regime of large complex structure (LCS) for some of the complex structure moduli. In particular, we consider two complex structure moduli which we label by $u$ and $v$. We then assume that the F-term conditions stabilise $u$ and $v$ such that the following hierarchy is realised:
\be
\label{eq:hierarchy1}
e^{-2 \pi \Im(u)} \ll e^{-2 \pi \Im(v)} \ll 1 \ .
\ee 
As we will show later, inflation proceeds along a direction in which the condition \eqref{eq:hierarchy1} remains true.
At LCS for $u$ and $v$ terms of the form $e^{2 \pi i u}$ and $e^{2 \pi iv}$ are suppressed and the leading term in the K\"ahler potential only depends on the shift-symmetric combinations $(u-\ub)$ and $(v - \vb)$. By enforcing \eqref{eq:hierarchy1} terms of the form $e^{2 \pi iu}$ are subleading compared to $e^{2 \pi iv}$ and are thus ignored in the following. We only retain the leading instantonic term in $v$. With the mild assumption that the dominant such term contributes as $e^{2 \pi iv}$ (i.e.~assuming $\beta_v=1$) we obtain the term $\mathcal{B} e^{2 \pi i v}$ in \eqref{eq:K1}.
The superpotential \eqref{eq:W1} is the Gukov-Vafa-Witten superpotential and takes the form 
\be
\W= (N_F- \tau N_H)^{\alpha} \Pi_{\alpha} \ ,
\ee
where $N_F^{\alpha}, N_H^{\alpha} \in \mathbb{Z}$ are flux integers, $\tau$ is the axio-dilaton and $\Pi$ is the period vector. It has entries
\begin{align}
& \Pi = \\
\nn & \begin{pmatrix}
  1 \\
 z^{i} \\
\frac{1}{2!} \kappa_{ijk} z^j z^k + \frac{1}{2!} a_{ij} z^j + b_{i} + \underset{{\beta, m}}{\sum} \frac{n_{\beta} \ \beta_{i} }{(2 \pi i m)^2} e^{2 \pi i m \beta_{i} z^{i}} \\
- \frac{\kappa_{ijk}}{3!} z^{i} z^j z^k + b_{i} z^{i} + \frac{c}{2} + \underset{{\beta, m}}{\sum} \frac{2 n_{\beta} (1-\pi i m \beta_{i} z^{i})}{(2 \pi i m)^3} e^{2 \pi i m \beta_{i} z^{i}}
 \end{pmatrix} .
\end{align}
Here the parameters $a_{ij}$ and $b_{i}$ also appear which we take to be real. 

The structure observed in \eqref{eq:W1} can be recovered as follows. By choosing appropriate flux integers $N_F^{\alpha}, N_H^{\alpha}$ one can ensure that $u$ and $v$ only appear linearly in $\W$ at leading order. In particular, the last entry of the period vector will typically give rise to quadratic and cubic terms in $u$ and $v$. We forbid these contributions by setting the last entry of the flux vector to zero. Choosing certain flux numbers large we can ensure that $N \gg1$. As before, instantonic terms in $u$ are ignored. We include the leading instantonic contribution in $v$ which gives rise to $g(z) e^{2 \pi i v}$ in \eqref{eq:W1}. 

In the following, we denote the remaining $(n-2)$ complex structure moduli (except $u$ and $v$) as well as the axio-dilaton by $z$. All in all, we find that the structure of \eqref{eq:K1} and \eqref{eq:W1} can indeed be realised in the complex structure moduli sector of a Calabi-Yau 3-fold. 

\subsection{Moduli stabilisation and Scalar potential}
Moduli stabilisation and the generation of the inflationary potential then proceed as follows. Besides fixing complex structure moduli and the axio-dilaton, we also need to stabilise K\"ahler moduli, which we do according to the Large Volume Scenario (LVS) \cite{0502058}. At leading order the theory for K\"ahler moduli is of no-scale type giving rise to the no-scale cancellation in $V$:
\begin{align}
\label{eq:scalpot}
\nn V &= e^{\K} \left(\K^{I \bar{J}} D_{I} \W \overline{D_{J} \W} + \K^{T_{\rho} \bar{T}_{\sigma}} D_{T_{\rho}} \W \overline{D_{T_{\sigma}} \W} - 3 {| \W|}^2 \right) \\
&= e^{\K} \K^{I \bar{J}} D_{I} \W \overline{D_{J} \W} \ ,
\end{align}
with $I,J$ running over all complex structure moduli and the axio-dilaton. The potential is thus minimised for $D_I \W=0$ for all $I$. Subleading terms due to $\alpha'$ and non-perturbative corrections stabilise the K\"ahler moduli.

We now focus on the complex structure moduli $u$ and $v$, in particular on their axionic directions Re$(u)$, Re($v$). Ignoring the exponential terms in \eqref{eq:K1} and \eqref{eq:W1}, only the combination Re($u$)$-N$Re($v$) appears in the scalar potential \eqref{eq:scalpot}, i.e. $V\equiv f\left[\Re(u)-N\Re(v)\right]$. Furthermore, the function $f$ is minimised at some argument $x_{0}$, i.e. at Re($u$)$-N$Re$(v)=x_{0}$. The latter equation defines a flat direction or  ``valley'' on the plane parameterised by Re($u$) and Re($v$). This flat direction is independent of the metric on field space, i.e. $\mathcal{K}$. It can be parameterised by any combination of Re$(u)$ and Re$(v)$, except the fixed combination Re($u$)$-N$Re($v$). In particular, we find it convenient to define the fields $\psi$ and $\phi$ by:
\be
\label{eq:basischange}
\phi \equiv u \ , \qquad \psi \equiv u - Nv \ .
\ee
With respect to these new variables, the flat direction is parameterised by Re$(\phi)$ with Re($\psi$) being fixed. Note that varying Re($\phi$) at fixed Re($\psi$) is \emph{not} the same as varying Re($u$) at fixed Re($v$), i.e.~the valley is not identical to the coordinate axis Re($u$).

Definitions of $\phi$ which differ from the one in \eqref{eq:basischange} are equally valid. One could e.g.~choose the new variable such that it describes a direction which is orthogonal to $\psi$ in field space. However, this is a metric dependent statement. Therefore, the definition of such orthogonal variables would involve elements of the K\" ahler metric and complicate the following computations. We prefer to redefine variables according to \eqref{eq:basischange}.

After the change of variables, the exponential term reads:
\begin{equation}
e^{-2\pi\Im(v)}=e^{-2\pi\frac{\Im(\phi)-\Im(\psi)}{N}}\equiv\epsilon\ll 1,
\end{equation}
where we defined the small parameter $\epsilon$ for notational convenience. To be precise, $\epsilon\equiv e^{-2\pi\Im(v_{0})}$, where $\Im(v_{0})$ is the value stabilised by the leading order (non-exponential) potential. As we will see, shifts in $v$ or, in the new variables, in $\psi, \phi$, will be small enough during inflation. Thus, $\epsilon$ is now a parameter.

In terms of our new variables the leading parts of $\K$ and $\W$ are
\begin{align}
\mathcal{K} &= K(z, \zb, \Im(\phi), \Im(\psi)) + \mathcal{O}(\epsilon) \ , \\
\W &= w(z) + f(z) \psi + \mathcal{O}(\epsilon) \ .
\end{align}
The conditions $D_I \W=0$ will in general stabilise all of the moduli $z$, both $\Im(\phi)$ and $\Im(\psi)$ as well as the combination $\Re(\psi)$.\footnote{Here $D_I \W=0$ for all $I$ gives rise to $2n+1$ real equations for $2n+2$ real moduli. Note that $D_{\phi} \W = K_{\phi} \W=0$ gives rise to only one real equation $K_{\phi}=0$.} In other words, the appearance of $\psi$ in $\W$ leads to the breakdown of the two shift symmetries in $u$ and $v$ to one remaining shift symmetry. This shift-symmetric direction is parametrised by $\Re(\phi)$, which does not appear in $\mathcal{K}$ and $\W$. It is our inflaton candidate, which has a flat potential at this stage. 

\begin{figure}
\centering
 \includegraphics[width=0.25\textwidth]{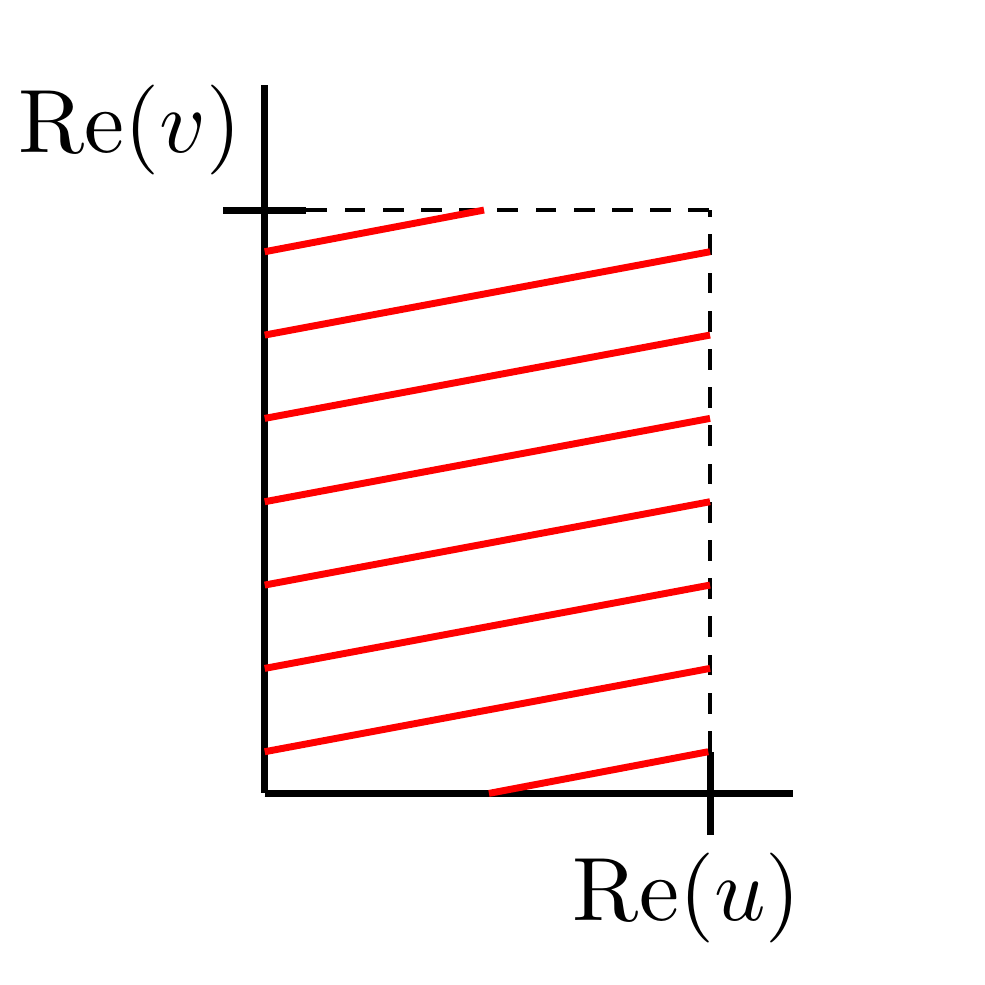}
\caption{Inflaton trajectory in $\textrm{Re}(v)$-$\textrm{Re}(u)$-plane. The winding trajectory is a result of stabilising one direction in $\textrm{Re}(v)$-$\textrm{Re}(u)$-space by an $F$-term potential due to bulk fluxes.}\label{fig:trajectory}
\end{figure}

It is easy to see from \eqref{eq:basischange} that, at fixed $\Re(\psi)$, the field $\Re(\phi)$ parametrizes a direction that is nearly aligned with $\Re(u)$. Indeed, we have $\delta\Re(u)=N\delta\Re(v)$ such that, at large $N$, $u$ changes much more strongly than $v$. Such a flat direction thus corresponds to a winding trajectory shown in figure \ref{fig:trajectory} and can be very long.

The idea of achieving a long direction due to winding trajectories in the field space of two axions was proposed in \cite{0409138} and subsequently many large-field inflation models in string theory have employed this mechanism \cite{09121341, 14047127, 14047773, 14047852, 14104660, 150301015, 150302965}. However, in contrast to previous proposals the winding trajectory in our case arises from an $F$-term potential due to fluxes. For example, in \cite{14047773} non-perturbative contributions to the potential were employed to realise a winding trajectory similar to ours (see also the discussion in Sec. \ref{sec:nogo}).
 
\subsection{Inflaton potential}
We now include the subleading terms $\sim e^{2\pi i v}= e^{2\pi i \frac{\phi-\psi}{N}} =\epsilon~e^{2\pi i\frac{\Re(\phi)-\Re(\psi)}{N}}$ and determine the inflaton potential to order $\epsilon^{2}$. We also include the backreaction on the remaining moduli in our analysis.

We can write $\K$ and $\W$ of \eqref{eq:K1} and \eqref{eq:W1} as
\begin{align}
\K &= K + \delta K + \mathcal{O}(\epsilon^2) \\
\W &= W + \delta W + \mathcal{O}(\epsilon^2) \ ,
\end{align}
where $\delta K \sim \delta W \sim \epsilon$. We thus have
\begin{align}
D_I \W = &\left(\partial_I W + K_I W \right) \\ 
\nn & + \left(\partial_I \delta W + K_I \delta W + \delta K_I W \right) + \mathcal{O}(\epsilon^2) \ ,
\end{align}
where the index $I$ runs over all superfields $z$, $\psi$ and $\phi$.
Given our specific structure for $\K$ and $\W$ from \eqref{eq:K1} and \eqref{eq:W1} we find that all covariant derivatives can always be brought into the form
\begin{align}
D_I \W = & A_I (z, \zb, \psi, \bar{\psi}, \phi-\bar{\phi}) \\
\nn & + \epsilon \left[B_I (z, \zb, \psi, \bar{\psi}, \phi-\bar{\phi})e^{\frac{2 \pi i \phi_1}{N}} \right. \\
\nn & \hphantom{+ \epsilon} \ \ \left. + C_I (z, \zb, \psi, \bar{\psi},\phi-\bar{\phi}) e^{-\frac{2 \pi i \phi_1}{N}} \right] \\
\nn & + \mathcal{O}(\epsilon^2) \ ,
\end{align}
where $\phi_1 \equiv \Re(\phi)$ and $A_I, B_I, C_I$ are complex functions of $z$, $\zb$, $\psi$, $\bar{\psi}$ and $\Im(\phi)$ which can be easily calculated for a specific example. Notice that we have reabsorbed the phases $e^{-2\pi i\psi_{1}/N}$ and $e^{2\pi i\psi_{1}/N}$ in the complex prefactors $B_{I}$ and $C_{I}$ respectively.

The potential is given by
\be
V=e^{K+\delta K} (K^{I \bar{J}} + \delta K^{I \bar{J}}) D_I \W \overline{D_{J} \W} + \ldots \ .
\ee
However, we will find that during inflation $D_{I} \W \sim \epsilon$. Thus, to determine $V$ at order $\epsilon^2$ we can ignore $\delta K$ and $ \delta K^{I \bar{J}}$ in the prefactor and only work with
\be
V=e^{K} K^{I \bar{J}} D_I \W \overline{D_{J} \W}  + \mathcal{O}(\epsilon^3) \ .
\ee
Without loss of generality we can use an orthogonal transformation $O_{I}^{J}$ to diagonalise the K\"ahler metric. Its eigenvalues $\lambda_{I}$ can then be reabsorbed in a redefinition of the rotated vectors $O_{I}^{J}(D_{J}\W)$.
The potential becomes 
\be
V=e^K \sum_{I=1}^{n+1} |v_I|^2 \ , \quad \textrm{with } v_I =  \sqrt{\lambda_I} \ \sum_J O_I^{\ J} (D_J \W) \ ,
\ee
where $\lambda_I$ are the eigenvalues of the original K\"ahler metric and $O$ is an orthogonal matrix which diagonalises the K\"ahler metric. Most importantly, $v_I$ will still have the same structure as $D_I \W$:
\be
v_I = \tilde{A}_I + \epsilon \left[\tilde{B}_I e^{\frac{2 \pi i \phi_1}{N}} +\tilde{C}_I e^{-\frac{2 \pi i \phi_1}{N}} \right] \ .
\ee
We then split $v_I$ into real and imaginary parts to write the potential as
\be
V=e^K \sum_{\alpha=1}^{2n+2} w_{\alpha}^2 \ ,
\ee
where 
\begin{equation}
 w_{\alpha} = \left\{
  \begin{array}{l l}
    \Re(v_{\alpha}) &\alpha=1, \ldots, n+1 \\
    \Im(v_{\alpha-(n+1)}) &\alpha=n+2, \ldots, 2n+2
  \end{array} \right. 
\end{equation} 
In particular, the $w_{\alpha}$ now have the structure
\be
w_{\alpha} = a_{\alpha}  + \epsilon \left[b_{\alpha} \cos  \left(\frac{2 \pi \phi_1}{N} \right) + c_{\alpha} \sin  \left(\frac{2 \pi \phi_1}{N} \right) \right] \ ,
\ee
where $a_{\alpha}$, $b_{\alpha}$ and $c_{\alpha}$ are functions of the $(2n+1)$ moduli $\Re(z^i), \Im(z^i), \Re(\psi), \Im(\psi), \Im(\phi)$, which we will denote by $\{ \xi^i \}$.

We now determine the inflaton potential including backreaction on the moduli $\{ \xi^i \}$. At leading order we found that the minimum of the potential was at $D_I W=0$ for all $I$. In terms of the real parameters this corresponds to $a_{\alpha}=0$ for all $\alpha$. Including the $\mathcal{O}(\epsilon)$ corrections giving rise to the inflaton potential, the minimum of the potential will now be shifted. The moduli $\{ \xi^i \}$ will be displaced from their values at the original SUSY minimum by a small amount $\delta \xi ^i$. We will find that this displacement is as small as $\delta \xi^i \sim \epsilon$. To determine the potential up to order $\epsilon^2$ it will thus be sufficient to expand the parameter $a_{\alpha}$ in $\delta \xi ^i$ about the SUSY minimum:
\be
a_{\alpha} = 0 + M_{\alpha i} \delta \xi^i + \mathcal{O}(\epsilon^2) \ .
\ee
We only need to keep the leading terms of $b_{\alpha}$ and $c_{\alpha}$ as any further expansion would only produce subleading terms.

Note that $a_{\alpha}$ is a vector with $(2n+2)$ entries which are linear combinations of $(2n+1)$ variables. This has the following consequences: we can always perform a rotation on $w_\alpha \rightarrow w_{\alpha}' = R w_\alpha$ such that $a_{\alpha}' =R a_{\alpha}$ is a vector whose last component is zero: $a_{2n+2}'=0$. We also change coordinates $\delta \xi^i \rightarrow (\delta \xi^i)' \equiv a_i^{\prime}$. The potential becomes
\begin{align}
V=& \ e^K \sum_{i=1}^{2n+1} \left\{ (\delta \xi^i)' + \epsilon \left[b_{i}^{\prime} \cos  \left(\frac{2 \pi \phi_1}{N} \right) \right. \right. \\
\nn & \hphantom{+AAAAAAAAAAA} \left. \left. + c_{i}^{\prime} \sin \left(\frac{2 \pi \phi_1}{N} \right) \right] \right\}^2 \\
\nn &+e^K \epsilon^2 \left(b_{2n+2}^{\prime} \cos  \left(\frac{2 \pi \phi_1}{N} \right) + c_{2n+2}^{\prime} \sin \left(\frac{2 \pi \phi_1}{N} \right)\right)^2 .
\end{align}
The effect of backreaction can be read off from the above expression. The moduli adjust such that $(\delta \xi^i)'$ cancel the first $(2n+1)$ contributions to the potential completely. The effective inflaton potential is then given by the last term alone. Dropping the indices on $b_{2n+2}^{\prime}$ and $c_{2n+2}^{\prime}$ we find:
\be
V_{\mathrm{inf}} = e^K \epsilon^2 \left(b^{\prime} \cos  \left(\frac{2 \pi \phi_1}{N} \right) +c^{\prime} \sin \left(\frac{2\pi \phi_1}{N} \right)\right)^2 \ ,
\ee
which one can rewrite as 
\be
\label{inflaton potential}
V_{\mathrm{inf}} = e^K \epsilon^2 \lambda^2 \left\{ \sin \left(\frac{2 \pi \phi_1}{N} + \theta \right) \right\}^2 \ .
\ee

\subsection{Compatibility with K\"ahler Moduli Stabilisation}

Finally, we analyse the compatibility of our inflation model with the LVS.  Phenomenologically, it is identical to natural inflation. Using the definitions of $u,v$ in terms of $\phi,\psi$ and $\partial\phi_{1}=N\partial\psi_{1}$ we have
\begin{equation}
\mathcal{L}\supset \mathcal{K}_{u\bar{u}}(\partial\phi_{1})^{2}+\mathcal{K}_{v\bar{v}}(\partial\phi_{1})^{2}/N^{2}+\mathcal{K}_{u\bar{v}}(\partial\phi_{1})^{2}/N+c.c.
\end{equation}
Therefore, for large $N$, we have $\K_{\phi\bar{\phi}}=\K_{u\bar{u}}+O(1/N)$ and the canonically normalised inflaton field is defined by: $\varphi=\sqrt{2\K_{u\bar{u}}}\phi_1+O(1/N)$. In terms of $\varphi$, the potential \eqref{inflaton potential} reads (after also using a trigonometric identity): 
\begin{equation}
V_{\mathrm{inf}}(\varphi) \sim  e^{K} \epsilon^2 \lambda^2 \left[ 1 - \cos \left(\frac{\varphi}{f}+2\theta \right) \right],
\end{equation} 
with the axion decay constant $f \sim N/(4\pi \mathrm{Im}(u)) $ and the canonically normalised inflaton $\varphi$.
Recent Planck results \cite{15020211} impose a lower bound of $f>6.8$ (at $95\%$ CL), and thus our model has to satisfy the phenomenological constraint 
\be \label{constraint on N}
\frac{N}{\mathrm{Im}(u)} > 85.
\ee
Notice that the flux number $N$ receives also an upper bound coming from the tadpole cancellation condition. If the tadpole $L_{\star} \equiv \chi (\mathrm{CY}_4)/24$ is not sufficiently large, it can turn out to be challenging to achieve $N \sim \mathcal{O}(100)$ to satisfy the above constraint. However, a detailed analysis could reveal that the above bound is too strict. So far we have not included backreaction of K\"ahler moduli which is expected to lead to a flattening of the inflaton potential \cite{10114521, 14112032} (see in particular the recent analysis of \cite{150105812}). In this case the bound on $N$ would have to be modified which we leave for future work. 

In the following, we demonstrate that the parameters $\lambda, \epsilon, \mathrm{Im}(u)$ and $N$ can be arranged such that the K\"ahler moduli can in principle be stabilised successfully in the framework of the LVS.  

In order to avoid destabilisation of the K\"ahler moduli during inflation, we require   
\begin{equation} \label{inflation constraint}
V_{\mathrm{inf}} \sim 10^{-8} \ll V_{\mathrm{LVS}} \sim e^K \frac{|W|^2}{\mathcal{V}^3} \sim \frac{g_s |W|^2}{\mathcal{V}^3 \mathrm{Im}(u)^3},
\end{equation}
where we used that the energy density of the inflaton potential in natural inflation is $V_{\mathrm{inf}} \sim 10^{-8}$ in Planck units.
Moreover, there is the well-known constraint on the ratio of the gravitino mass and the Kaluza-Klein scale \cite{13106694}: 
\begin{equation}
\frac{m_{3/2}}{m_{\mathrm{KK}}} \ll 1.
\end{equation}
This constraint can be translated into the condition\footnote{One can derive this relation as follows: the gravitino mass is given by $m_{3/2} \sim e^{\mathcal{K}/2}|W| \sim \sqrt{g_s}|W|\mathcal{V}^{-1} \mathrm{Im}(u)^{-3/2}$. In anisotropic compactifications, e.g. in cases where one modulus is stabilised in the large complex structure limit, the Kaluza-Klein mass scales as $m_{\mathrm{KK}} \sim L^{-1} \mathcal{V}^{-1/2} \sim (\mathrm{Im}(u))^{-1/2} \mathcal{V}^{-2/3}$, where we parametrised the volume as $\mathcal{V} \sim L^3R^3$ with $L \gg R$ and $\mathrm{Im}(u) \sim L/R$.} 
\begin{equation} \label{gravitino constraint}
	\frac{m_{3/2}}{m_{\mathrm{KK}}} \sim \frac{\sqrt{g_s}|W|}{\mathrm{Im}(u) \mathcal{V}^{1/3}} \ll 1.
\end{equation}
Combining the constraints \eqref{inflation constraint} and \eqref{gravitino constraint} yields 
\begin{equation} \label{combined constraint}
	10^{-8} \ll \frac{1}{\mathrm{Im}(u) \mathcal{V}^{7/3}}.
\end{equation}
In order to fulfil the bounds described above, we choose the following numerical example: $N=150$, $\mathrm{Im}(u) =1.5$, $\epsilon = 0.02$ (i.e.~$\mathrm{Im}(v)\simeq 0.6)$, $\mathcal{V}=100$ and $\lambda \sim \mathcal{O}(10)$. The choices for $\Im(u)$ and $\Im(v)$ ensure that terms of the form $e^{-2 \pi \Im(v)}$ are small while the terms $e^{-2 \pi \Im(u)}$ are negligible. At the same time it is ensured that inflation occurs at the correct energy scale $V_{\mathrm{inf}} \sim10^{-8}$.

In this case,  \eqref{constraint on N} and \eqref{combined constraint} are easily satisfied. If $g_s \sim 0.01$ and $W \lesssim \mathcal{O}(10)$, then \eqref{gravitino constraint} holds true as well. For this, we have to compensate the large contribution coming from $N=150$ by a mild tuning of the coefficient $f(z) \sim \mathcal{O}(0.1)$. Since the LVS potential is $\mathcal{O}(10)$ times larger than the inflaton potential, there is no danger of destabilisation of the K\"ahler moduli.

\section{Relation to No-Go Theorems}
\label{sec:nogo}
We now turn to `no-go theorems' against large-field inflation (and large-field axion models in particular), see e.g.~\cite{0601001, 12035476, 14095793, 150300795, 150303886, 150304783}. We want to make a few observations, triggered in particular by the very recent discussions in \cite{150303886, 150304783}, with the aim of assessing to which extent our specific model has a chance to survive closer scrutiny.

To begin, a rather general criticism \cite{150303886} of models with large axion decay constants can be made on the basis of 4d gravitational instantons \cite{Giddings:1987cg, 9604038, 9608065}. Their euclidean action is $S_E\sim n/f$ with $n$ the instanton number and $f$ the periodicity of the relevant axion. The maximal scalar curvature of the solution is $R_{max}\sim (f/n)^2$, which should not exceed the square of the UV cutoff, $\Lambda^2$, to be able to trust the Einstein-Hilbert action with a single scalar. Hence the lowest $n$ we can trust in the instanton sum is determined by $f/n \sim \Lambda$, and the instanton-induced correction to the potential is 
\be
\delta V \sim e^{-S_E} \sim e^{-n/f} \sim e^{-1/\Lambda}\,.
\ee
Note that here and below we ignore all numerical factors such as $(2\pi)$ etc.

The relative correction to the inflationary potential is then 
\be
\frac{\delta V}{V}\sim \frac{e^{-1/\Lambda}}{H^2}\,,
\ee
which is clearly disastrous for a Planck-scale cutoff $\Lambda\sim 1$ (and for $H\lesssim 1$ as required by consistency). 

However, if $\Lambda$ takes the smallest value consistent with an (at least approximate) description by single-field inflation, i.e.~$\Lambda\sim H$ , one has
\be
\frac{\delta V}{V}\sim \frac{e^{-1/H}}{H^2}\,,
\ee
which goes to zero very quickly as $H$ falls below the Planck scale. Now, in our specific model (as in many other attempts to realize inflation with moduli stabilization in string theory) one is forced into precisely this regime: A relatively large CY volume is needed for parametric control and, as a result, moduli tend to become light. Thus, we may for a fundamental reason be in the regime where the 4d gravitational instanton argument does not work straightforwardly.

To be very clear, we take this general no-go argument extremely seriously. However, we are at the moment unable to explicitly `kill' our construction with this argument. 

Next, it is argued in \cite{150303886} that stringy instantons are the natural `UV continuation' of the above 4d effect below $\Lambda$. Furthermore, in the presence of strong SUSY breaking, instantons should very generically contribute to the scalar potential \cite{08082918}, again destroying the flat axion potential. 

Now, part of such instantons (those of type $\exp(2\pi i u)$) are argued to be systematically suppressed. This is in fact a central part of our suggestion. However, we are certainly aware that the `genericity' argument of \cite{150303886} can nevertheless affect us in various ways. Two obvious potential problems are the following:

The familiar `small 4-cycle' instanton of the LVS construction, $A\exp(-2\pi\tau_s$), has a complex-structure dependent prefactor $A(z,u,v)$. We hope, based on the fact that this instanton is a highly local effect (associated with the small blow-up cycle), that there are models where the $u$-dependence of $A$ is subdominant. However, we do not know whether this can be realised. The dependence on $\Re(u)$ introduced by $A$ would be periodic and, structurally, a loop-effect (a loop-correction to the gauge kinetic-function for a brane stack wrapping the small 4-cycle). It might nevertheless fall into the category of `generic instanton corrections' above since, clearly, an open-string loop may be viewed as an open-string worldsheet instanton in the limit that the worldsheet can be shrunk to zero volume. 

Similarly, there are loop corrections to the K\"ahler-moduli K\"ahler potential which, however, affect the scalar potential only at subleading order \cite{0507131, 0508043, 08051029}. As the explicit string-loop results of \cite{0508043} demonstrate, these effects depend in general on complex-structure moduli and can ruin the inflationary potential in principle. One may however hope that (as shown in \cite{Neuenfeld} in a special case) the dependence on certain complex structure moduli is exponentially suppressed in the LCS limit. Furthermore, the suppression by the (volume)$^{1/3}$ factor relative to the leading-order LVS potential may be sufficient to control the effect.

We have to leave the detailed study of the above corrections to future work. 

We now turn to the weak gravity conjecture (WGC) \cite{0601001} or, more specifically, to the very interesting direct relation to models of axion inflation proposed in \cite{150304783}. First, we recall the main point of \cite{150304783} relevant for our purposes: Let's assume one has some type-II CY compactification leading to 4d $C$-form axions and instantons. The 4d axionic scalars arise from integrating the $p$-form $C_p$ over $p$-cycles and the instantons from wrapping euclidean $p-1$ branes on those cycles. Now, one of the 4 non-compact dimensions is compactified on $S^1$, T-dualized, and the dual radius is taken to infinity. In this process, the brane acquires an extra non-compact dimension, consistent with the change between IIA and IIB (or vice versa). The $p$-form acquires an extra index turning into a 4d vector. As a result, one is now in a setting with 4d vectors and charged particles, where the WGC applies. Its constraints on masses of particles and their charges under the vectors can then be ``pulled back'' to the actions of instantons and their couplings to the axions (i.e.~the axion decay constants).

In our case, however, the ``axion'' on the type IIB side is a metric fluctuation in 10 dimensions. It is unclear why a T-duality along the `external' $S^1$ direction would turn this into a vector. Furthermore, the ``instanton'' in our case is a purely geometric effect. It is equally difficult to understand how this could become a ``particle'' under a T-duality transformation. On the other hand, one may hope to connect to the arguments presented in \cite{150304783} by reinterpreting our model as the mirror-dual type IIA setup. Clearly, this only works `in spirit' rather than explicitly since the T-dual radii would be small on the IIA side and since some of the IIB fluxes will dualize to a non-Calabi-Yau geometry. Nevertheless, it is instructive to proceed. On the type IIA side, the axions descend from $B_2$ on CY 2-cycles while the instantons are worldsheet instantons wrapping these 2-cycles. Thus, even in type IIA the T-duality argument does not work straightforwardly: The fundamental string does not develop an extra dimension under T-duality in the `external' $S^1$ and $B_2$ does not acquire an extra index. 

However, we can relate the type-IIA interpretation to the WGC much more straightforwardly: Viewing type IIA as M-theory on $S^1$, we can understand $B_2$ as coming from $A_3$ of M-theory and the instanton as an M2-brane wrapping both the CY 2-cycle and the $S^1$. This allows us to use the WGC to constrain our type IIB (complex-structure) axions. However, we already know that these axions (on a CY without fluxes) will be sub-Planckian. The whole point of our concrete model is to make one of them effectively super-Planckian by the flux-induced $F$-term constraint. The non-trivial task would then be to develop the above M-theory picture for a model with orientifold planes, branes and fluxes on the IIB side. The IIA side would then also have an orientifold plane, branes and fluxes plus, in addition, non-CY geometry. At the very least, the M-theory product space $\textrm{CY} \times S^1 \times \mathbb{R}^4$ would then turn into a non-trivial, degenerate $S^1$ fibration. It again becomes unclear whether a constraint on our model from the WGC can be derived. 

Thus, to the best of our understanding, we cannot directly derive a WGC-based constraint of the effective, super-Planckian instanton of our model. Nevertheless, let us now suppose that the difficulties outlined above can be overcome and the constraint holds. As explained in \cite{150304783}, the implications of such a constraint for inflation then depend crucially on whether the strong or the mild version of the WGC is used. 
\begin{itemize}
\item If the strong version of the WGC is correct, all models with large axion decay constants in a calculable regime should be inconsistent.\footnote{By calculable regime we mean situations where $S>1$ and hence $\exp(-S)$ is small.} In this case one should be able to rule out our proposal: by studying our model in more detail we would then expect obstacles to large field inflation to appear. 
\item On the other hand, if the WGC only holds in its mild form, there is a loophole \cite{150300795, 150304783} which allows for simultaneously satisfying the mild WGC and realising transplanckian axion inflation. Very intriguingly, our model appears to fit precisely into this loophole.
\end{itemize}

At this point, it is important to notice that there exists yet another loophole in the no-go arguments based on the WGC \cite{14123457, 150300795}. This loophole exploits the fact that the WGC for electrically charged objects can be satisfied for $f > 1$ if $S$ is taken to be sufficiently small, i.e.~$S \ll 1$. However, for $S<1$ we can in general not calculate the instanton induced potential and it hence remains unclear whether inflation can be realized. In concrete models, the potential can be calculable in spite of $S<1$ and inflation might still work. Such a possibility was suggested in \cite{14123457} (see also \cite{150300795}), based on the model of \cite{14056720}.\footnote{Consistency with the magnetic form of the WGC constrains models exploiting the loophole of \cite{14123457, 150300795} even further. As was pointed out in \cite{14123457} a transplanckian field range can only be achieved in a field space of two or more axions, e.g.~by employing a mechanism like alignment \cite{14056720}.}

Let us return to the loophole of \cite{150300795, 150304783}. To be consistent with the mild form of the WGC the relevant axion $\phi$ has to couple to two instantons giving rise to a potential of the form
\begin{align}
\label{eq:looph}
V = \Lambda_1^4 e^{-m} \left[1 -\cos \left(\frac{\phi}{f} \right) \right] + \Lambda_2^4 e^{-M} \left[1 -\cos \left(\frac{k \phi}{f} \right) \right] \ ,
\end{align}
with $k \in \mathbb{Z}$. As long as $f/k$ is subplanckian the mild form of the WGC is satisfied in virtue of the second term, even for a transplanckian $f$. The second term is not suitable for inflation while the first term can sustain inflation for transplanckian $f$. A successful model for inflation can then be achieved when $e^{-M}\ll e^{-m}$ such that contributions from the second term are suppressed.

We observe this structure in our setup. We begin with two instantons $e^{2 \pi i u}$ and $e^{2 \pi i v}$. We stabilise and integrate out the ``saxions'' $u_2 \equiv \Im(u)$, $v_2 \equiv \Im(v)$ such that $\exp(-2 \pi u_2) \ll \exp(-2 \pi v_2)$. In addition, we stabilise one axionic direction $\Re(\psi)=\Re(u) - N \Re(v)$ such that we are left with one axion $\phi_1=\Re(u)$ coupling to two instantons. Canonically normalising $\varphi= \sqrt{2K_{\phi \bar{\phi}}} \ \phi_1$ and defining $f_{eff}= N \sqrt{2 K_{\phi \bar{\phi}}}\,$, the axion $\varphi$ couples to instantonic terms as 
\be
\label{eq:instantons}
A_1e^{-2 \pi v_2} e^{2 \pi i \frac{\varphi}{f_{eff}}} \quad \textrm{and} \quad A_2 e^{-2 \pi u_2} e^{2 \pi i \frac{ N \varphi}{f_{eff}}} \ .
\ee
This is just as in the loophole presented above. Notice that we have here performed the change of basis only for the real parts of $u$ and $v$ , to make connection with \eqref{eq:looph}. Choosing $N \gg 1$ one can achieve a transplanckian $f_{eff}$. The first term in \eqref{eq:instantons} then gives rise to the inflaton potential, while the second instanton ensures that the mild form of the WGC is satisfied. By having stabilised $\exp(-2 \pi u_2) \ll \exp(-2 \pi v_2)$ we also prevent the second term from spoiling the inflaton potential.

This feature distinguishes our model from previous proposals, such as \cite{14047773}. The latter model realises axion alignment with two periodic terms and two axions. Therefore, as explained in \cite{150300795}, it violates the WGC. Moreover, the authors employ a superpotential of the form:
\begin{equation}
\W_{inf}=A_{1}e^{-\frac{2\pi\psi}{f_{1}}}e^{-\frac{2\pi\phi}{f_{2}}}+A_{2}e^{-\frac{2\pi\psi}{f_{3}}},
\end{equation}
with $f_{2}$ superplanckian and $f_{1},f_{3}$ subplanckian. In their model, the imaginary part of $\phi$ is the inflaton candidate. It appears only in one term in the superpotential, with a large decay constant. Therefore, there seems to be no other term in $\phi_{1}$ which satisfies the WGC and which is more suppressed than the inflationary one. Note that in principle one can enforce the WGC by adding other non-perturbative terms in $\phi$, and making sure that a hierarchy in the decay constants can still be realised. 

The fact that our setup exhibits the right properties to realise the loophole of \cite{150300795, 150304783} heavily relies on the fact that we can stabilise the saxions $u_2$ and $v_2$ and one combination of the axions $u_1$ and $u_2$ independently. Any further detailed analysis of our setup should thus focus on whether the required pattern of moduli stabilisation can be achieved in an explicit example. In fact, such further scrutiny could indeed be very rewarding: if our model was confirmed to be consistent, this would correspond to evidence that the WGC holds in string theory only in its mild form.

\section{Conclusions}
We introduced a new model of large field inflation in string theory, which employs axionic fields arising from complex structure moduli of a Calabi-Yau 3-fold at large complex structure. The transplanckian field range is generated as a winding trajectory in the compact field space of two (or more) axions. One new aspect of this construction is that this winding trajectory can be generated by an $F$-term potential from bulk fluxes in type IIB string theory. Such a trajectory requires a hierarchy in the flux superpotential, which can be achieved by choosing a set of flux numbers large. As the size of flux numbers is limited by tadpole cancellation there is a natural cutoff for the axion field range, such that it cannot get arbitrarily large.

The inflaton potential is generated by corrections to the geometry, which take an ``instantonic'' oscillatory form and give rise to a model of natural inflation. The smallness of the instantonic correction relieves some tension between inflation and K\"ahler moduli stabilisation, as has been observed in axion monodromy models in the complex structure moduli sector \cite{14112032}(see however \cite{150301607}). In particular, we can make the inflaton potential subdominant without further excessive tuning.

Clearly, there is further work needed on the model presented here. In particular, it would be important to verify in a concrete example that the proposed pattern for moduli stabilisation can indeed be achieved. Further, while we included the backreaction of complex structure moduli, a similar analysis for K\"ahler moduli along the lines of \cite{150105812} would be desirable. We also only presented a very preliminary analysis of the phenomenology of the resulting inflaton potential.

Our model realises axion inflation with an effective transplanckian axion period and is thus potentially in conflict with the recent ``no-go theorems''. However, while these analyses are extremely important for this field of study, there are still many open questions how these results apply to concrete models. In particular, we find it is challenging to translate the results in \cite{150304783} -- which are derived for $C_2$ axions -- to our construction, which employs axionic directions in complex structure moduli space. Further no-go results based on gravitational instantons \cite{150303886} only become constraining in a regime where classical gravity cannot be trusted. 

However, even if no-go arguments based on applying the weak gravity conjecture to axions hold, our construction appears to realise a loophole pointed out in \cite{150300795,150304783}. Given the current theoretical understanding, we thus believe that our construction is a valuable addition to the set of large field models in string theory. 

\subsection*{Acknowledgments}
We would like to thank Jose Francisco Morales and Timo Weigand for insightful discussions. We also thank Renata Kallosh and Timm Wrase for comments on moduli stabilisation. This work was partly supported by the DFG Transregional Collaborative Research Centre TRR 33 ``The Dark Universe''. During the completion of this work, P.M.~and F.R.~were partially supported by DFG Graduiertenkolleg GRK 1940 ``Physics Beyond
the SM''. Moreover, F.R.~was supported by the DAAD. P.M.~also acknowledges support from the Studienstiftung des deutschen Volkes.

\bibliography{windingbib}
\end{document}